\documentclass[]{raa}
\usepackage{graphicx,times}
\usepackage{natbib}
\usepackage{amssymb,amsmath}
\usepackage{longtable}
\usepackage{subfigure}

\begin{document}
\title{Long-term variation of the solar polar magnetic fields at different latitudes}

 \volnopage{ {\bf 2024} Vol.\ {\bf 24} No. {\bf 7}, 075015}
   \setcounter{page}{1}

\author{Shuhong Yang\inst{1,2,7}, Jie Jiang\inst{3}, Zifan Wang\inst{1,2,7}, Yijun Hou\inst{1,2,7}, Chunlan Jin\inst{1,2,7}, Qiao Song\inst{4,6}, Yukun Luo\inst{3}, Ting Li\inst{1,2,7}, Jun Zhang\inst{5}, Yuzong Zhang\inst{1,2,7}, Guiping Zhou\inst{1,2,7}, Yuanyong Deng\inst{1,2,7}, Jingxiu Wang\inst{1,2,7}}

\institute{$^1$National Astronomical Observatories, Chinese Academy of Sciences, Beijing 100101, China; {\it shuhongyang@nao.cas.cn} \\
           $^2$School of Astronomy and Space Science, University of Chinese Academy of Sciences, Beijing 100049, China \\
           $^3$School of Space and Environment, Beihang University, Beijing 102206, China; {\it jiejiang@buaa.edu.cn} \\
           $^4$Key Laboratory of Space Weather, National Satellite Meteorological Center (National Center for Space Weather), China Meteorological Administration, Beijing 100081, China \\
           $^5$School of Physics and Materials Science, Anhui University, Hefei 230601, China \\
           $^6$Innovation Center for FengYun Meteorological Satellite (FYSIC), Beijing 100081, China \\
           $^7$State Key Laboratory of Solar Activity and Space Weather, National Space Science Center, Chinese Academy of Sciences, Beijing 100190, China \\
\vs\no
   {\small \it Received 2024 April 11; revised 2024 May 28; accepted 2024 May 31; published 2024 July 3 }}

\abstract{
The polar magnetic fields of the Sun play an important role in governing solar activity and powering fast solar wind. However, because our view of the Sun is limited in the ecliptic plane, the polar regions remain largely uncharted. Using the high spatial resolution and polarimetric precision vector magnetograms observed by Hinode from 2012 to 2021, we investigate the long-term variation of the magnetic fields in polar caps at different latitudes. The Hinode magnetic measurements show that the polarity reversal processes in the north and south polar caps are non-simultaneous. The variation of the averaged radial magnetic flux density reveals that, in each polar cap, the polarity reversal is completed successively from the 70$\degr$ latitude to the pole, reflecting a poleward magnetic flux migration therein.
These results clarify the polar magnetic polarity reversal process at different latitudes.
}

\authorrunning{Yang et al.}
\titlerunning{Solar polar magnetic fields}
\maketitle

\section{Introduction}

Exploring the solar poles is a great frontier of the Sun and is vital to understand the driver of long-term solar cycle, short-term solar activity, fast solar wind, space weather, and also stellar cycle (Charbonneau 2020). As the primary determinant, the Sun's polar magnetic fields are considered to be a direct manifestation of the solar interior and serve as seed fields for the global dynamo producing the solar cycle (Cameron et al. 2016). Due to the location in the plane of the ecliptic, there are large projection effects when we observe the solar polar regions from/near the Earth. This limits the precise measurement of the magnetic fields in the high latitude region of the Sun.

Since the polar magnetic fields are dominated by small-scale magnetic flux concentrations, it is an outstanding challenge to accurately measure the polar magnetic fields.
The pioneering observations of the polar regions only provide the longitudinal magnetic field measurement, and the polar magnetic fields are assumed to be approximately radial (Tang \& Wang 1991; Lin et al. 1994; Homann et al. 1997). The vector magnetic field measurement can obtain the detailed information about polar magnetic structure.
The vector magnetic field of the solar polar region was first systematically measured with a ground-based instrument in 1997 (Deng et al. 1999).
However, it is hard for ground-based telescopes to provide an accurate measurement of the magnetic fields near the solar poles, because of the existence of atmospheric seeing.
Therefore, we are lacking knowledge about the long-term variation and spatial distribution of the precise magnetic fields in the polar caps.

As a space mission, Hinode (Kosugi et al. 2007) satellite provides high spatial resolution and polarimetric precision observations of the polar vector magnetic fields (Tsuneta et al. 2008a; Ito et al. 2010; Jin \& Wang 2011; Shiota et al. 2012). The Hinode has routinely measured the polar vector magnetic fields over a solar cycle, thus giving us a unique opportunity to investigate the long-term variation of the polar magnetic fields.
Using the ten-year vector magnetic field observations from 2012 to 2021 carried out with the Spectro-polarimeter (SP; Lites et al. 2013) of the Solar Optical Telescope (SOT; Tsuneta et al. 2008b) aboard Hinode, we study the long-term variation of the solar polar magnetic fields at different latitudes.

\section{Observations and Results}

As one of the most important facilities onboard Hinode, the SOT/SP measures the vector magnetic fields in the photosphere with high spatial and spectral resolutions.
We use the Hinode/SP polar observations\footnote[1]{https://hinode.isee.nagoya-u.ac.jp/sot\_polar\_field/} released at ISEE, Nagoya University. The Hinode data adopted in this study are listed in Table \ref{tbl}.
The pixel size in the east-west direction (i.e., the SP scanning direction) is approximately 0.30{\arcsec} and that in the north-south direction (i.e., along the slit) is almost 0.32{\arcsec}.
The magnetic field data were retrieved from the SP full stokes profiles of Fe~{\sc{i}} 630.15 nm and 630.25 nm by applying the Milne-Eddington inversion (Orozco Su{\'a}rez \& Del Toro Iniesta 2007).
Then the retrieved magnetic data were further processed to resolve the 180$^\circ$ ambiguity (Ito et al. 2010) and obtain the radial magnetic field component (Shiota et al. 2012).

\begin{longtable}{ccccccc}
\caption{Hinode SOT/SP Polar Observations} \label{tbl}\\
\hline\hline
\multicolumn{3}{c}{South} & & \multicolumn{3}{c}{North} \\
\cline{1-3}  \cline{5-7}
Date & \multicolumn{2}{c}{Time (UT)} & & Date & \multicolumn{2}{c}{Time (UT)} \\
\cline{2-3}  \cline{6-7}
 & Start & End & & & Start & End \\
\hline
2012-04-03 & 11:10:56 & 13:59:45 & & 2012-09-10 & 09:05:05 & 11:58:39 \\
2012-04-06 & 09:35:23 & 12:24:53 & & 2012-09-14 & 10:05:05 & 12:58:19 \\
2012-04-09 & 12:05:22 & 14:58:56 & & 2012-09-16 & 10:45:05 & 13:38:38 \\
2012-04-12 & 09:55:24 & 12:48:58 & & 2012-09-19 & 10:06:56 & 12:54:53 \\
2012-04-15 & 10:05:23 & 12:58:57 & & 2012-09-22 & 09:56:06 & 12:49:39 \\
2012-04-19 & 11:18:23 & 14:11:57 & & 2012-09-25 & 23:04:23 & +1 01:57:57 \\
2012-04-21 & 10:55:24 & 13:48:56 & & 2012-09-28 & 23:04:22 & +1 01:57:56 \\
2012-04-24 & 11:08:23 & 14:01:56 & & 2012-10-01 & 10:09:59 & 13:03:33 \\
2012-04-27 & 10:06:06 & 12:59:39 & & 2012-10-04 & 10:58:22 & 13:51:55 \\
2012-04-30 & 10:06:06 & 12:59:39 & & 2012-10-07 & 12:05:22 & 14:54:51 \\
\hline
2013-03-01 & 10:06:05 & 12:54:53 & & 2013-09-02 & 09:05:44 & 11:59:18 \\
2013-03-04 & 10:40:34 & 13:34:07 & & 2013-09-04 & 09:40:44 & 12:33:47 \\
2013-03-07 & 09:42:05 & 12:35:38 & & 2013-09-06 & 10:41:29 & 13:29:47 \\
2013-03-10 & 09:36:00 & 12:29:33 & & 2013-09-09 & 10:50:05 & 13:43:39 \\
2013-03-13 & 10:06:00 & 12:59:34 & & 2013-09-13 & 10:31:46 & 13:24:48 \\
2013-03-16 & 10:53:00 & 13:43:51 & & 2013-09-16 & 10:35:46 & 13:29:08 \\
2013-03-19 & 10:35:22 & 13:28:55 & & 2013-09-19 & 10:57:01 & 13:46:50 \\
2013-03-22 & 10:06:37 & 13:00:10 & & 2013-09-22 & 10:52:08 & 13:27:02 \\
2013-03-25 & 10:07:37 & 12:59:49 & & 2013-09-25 & 10:07:09 & 12:59:52 \\
2013-03-28 & 09:36:46 & 12:30:19 & & 2013-09-28 & 10:27:23 & 13:20:58 \\
\hline
2014-02-28 & 17:04:06 & 19:57:39 & & 2014-08-28 & 10:29:09 & 13:22:41 \\
2014-03-03 & 10:05:34 & 12:59:08 & & 2014-08-31 & 11:07:09 & 14:00:43 \\
2014-03-06 & 10:08:35 & 12:59:46 & & 2014-09-03 & 11:07:12 & 14:00:45 \\
2014-03-09 & 11:01:37 & 13:55:10 & & 2014-09-09 & 11:14:37 & 14:04:47 \\
2014-03-12 & 15:05:34 & 17:59:08 & & 2014-09-13 & 20:24:23 & 23:17:56 \\
2014-03-15 & 18:15:35 & 21:09:08 & & 2014-09-15 & 22:51:05 & +1 01:44:39 \\
2014-03-18 & 09:50:38 & 12:44:11 & & 2014-09-18 & 21:46:23 & +1 00:39:56 \\
2014-03-21 & 15:16:14 & 18:09:47 & & 2014-09-21 & 11:07:09 & 13:59:51 \\
2014-03-24 & 08:15:23 & 11:08:56 & & 2014-09-24 & 11:06:01 & 13:59:34 \\
2014-03-27 & 18:47:51 & 21:33:26 & & & & \\
\hline
2015-02-24 & 11:00:23 & 13:53:57 & & 2015-08-30 & 10:07:11 & 13:00:45 \\
2015-02-27 & 12:05:23 & 14:58:57 & & 2015-09-02 & 11:35:21 & 14:28:55 \\
2015-03-02 & 10:45:20 & 13:38:54 & & 2015-09-05 & 11:05:35 & 13:59:08 \\
2015-03-05 & 10:18:10 & 13:11:43 & & 2015-09-08 & 11:06:22 & 13:54:49 \\
2015-03-08 & 09:05:38 & 11:59:11 & & 2015-09-11 & 11:07:12 & 13:54:48 \\
2015-03-11 & 09:06:07 & 11:59:40 & & 2015-09-14 & 10:06:06 & 12:59:39 \\
2015-03-14 & 10:45:09 & 13:38:41 & & 2015-09-17 & 12:50:22 & 15:43:56 \\
2015-03-17 & 10:39:07 & 13:32:41 & & 2015-09-20 & 10:35:23 & 13:28:56 \\
2015-03-19 & 10:27:22 & 13:20:56 & & 2015-09-23 & 10:36:52 & 13:30:26 \\
2015-03-23 & 10:19:22 & 13:12:56 & & 2015-09-26 & 10:52:05 & 13:45:38 \\
            &           &        & & 2015-09-29 & 11:03:09 & 13:56:43 \\
\hline
2016-03-03 & 10:55:08 & 13:48:41 & & 2016-08-26 & 06:19:56 & 09:13:29 \\
2016-03-05 & 10:28:06 & 13:21:39 & & 2016-08-29 & 10:06:04 & 12:59:37 \\
2016-03-08 & 10:04:31 & 12:58:04 & & 2016-09-01 & 12:47:06 & 15:40:40 \\
2016-03-11 & 10:05:22 & 12:58:54 & & 2016-09-04 & 09:55:29 & 12:49:02 \\
2016-03-14 & 10:06:04 & 12:59:36 & & 2016-09-08 & 11:30:31 & 14:24:03 \\
2016-03-17 & 11:02:05 & 13:55:39 & & 2016-09-10 & 12:45:07 & 15:38:40 \\
2016-03-20 & 10:18:22 & 13:11:55 & & 2016-09-13 & 10:40:07 & 13:33:41 \\
2016-03-23 & 12:51:04 & 15:40:54 & & 2016-09-16 & 10:30:56 & 13:24:30 \\
2016-03-26 & 10:23:08 & 13:16:42 & & 2016-09-19 & 10:35:43 & 13:29:16 \\
2016-03-29 & 10:43:28 & 13:37:01 & & 2016-09-22 & 17:15:45 & 20:09:18 \\
2016-04-01 & 10:06:11 & 12:59:45 & & & & \\
\hline
2017-02-22 & 10:15:27 & 13:05:46 & & 2017-08-23 & 10:35:06 & 13:28:40 \\
2017-02-25 & 10:30:06 & 13:23:39 & & 2017-08-26 & 11:31:08 & 14:24:40 \\
2017-02-28 & 10:15:53 & 13:00:36 & & 2017-08-29 & 11:36:06 & 14:29:39 \\
2017-03-03 & 10:05:45 & 12:59:18 & & 2017-09-01 & 10:51:07 & 13:40:46 \\
2017-03-06 & 10:05:06 & 12:58:39 & & 2017-09-04 & 09:25:06 & 12:18:39 \\
2017-03-09 & 10:30:40 & 13:24:13 & & 2017-09-13 & 10:05:38 & 12:59:12 \\
2017-03-12 & 00:06:06 & 02:59:39 & & 2017-09-16 & 10:49:03 & 13:42:36 \\
2017-03-15 & 12:55:07 & 15:48:40 & & 2017-09-19 & 10:25:02 & 13:18:36 \\
2017-03-18 & 10:35:02 & 13:28:36 & & 2017-09-22 & 13:06:08 & 15:59:42 \\
2017-03-22 & 05:48:08 & 08:41:42 & & & & \\
\hline
2018-02-22 & 11:12:08 & 14:05:41 & & 2018-08-24 & 10:08:18 & 13:01:52 \\
2018-02-25 & 07:16:21 & 10:09:54 & & 2018-08-27 & 11:46:06 & 14:39:39 \\
2018-02-27 & 09:47:21 & 12:39:53 & & 2018-08-30 & 20:36:02 & 23:29:36 \\
2018-03-03 & 11:02:22 & 13:55:55 & & 2018-09-02 & 10:20:36 & 13:14:09 \\
2018-03-06 & 11:11:08 & 14:04:41 & & 2018-09-05 & 11:34:37 & 14:28:11 \\
2018-03-08 & 11:14:47 & 14:08:21 & & 2018-09-11 & 14:20:06 & 17:13:40 \\
2018-03-12 & 11:01:21 & 13:31:29 & & 2018-09-14 & 18:13:36 & 21:07:10 \\
2018-03-15 & 23:12:00 & +1 02:05:33 & & 2018-09-17 & 11:04:18 & 13:57:40 \\
2018-03-18 & 09:50:22 & 12:43:46 & & 2018-09-20 & 10:55:06 & 13:48:39 \\
2018-03-24 & 10:52:03 & 13:45:36 & & 2018-09-23 & 09:39:37 & 12:33:11 \\
\hline
2019-02-20 & 10:05:22 & 12:58:56 & & 2019-08-25 & 10:06:13 & 12:54:50 \\
2019-02-23 & 11:10:48 & 14:04:21 & & 2019-08-28 & 10:06:13 & 12:54:51 \\
2019-02-26 & 11:30:23 & 14:23:57 & & 2019-08-31 & 10:50:16 & 13:43:49 \\
2019-03-01 & 09:52:51 & 12:46:25 & & 2019-09-03 & 16:01:22 & 18:54:55 \\
2019-03-04 & 10:00:31 & 12:54:03 & & 2019-09-06 & 10:05:21 & 12:58:55 \\
2019-03-08 & 10:39:46 & 13:33:20 & & 2019-09-09 & 08:35:21 & 11:28:54 \\
2019-03-10 & 10:13:46 & 13:07:18 & & 2019-09-12 & 08:02:06 & 10:55:40 \\
2019-03-13 & 11:05:21 & 13:58:55 & & 2019-09-15 & 10:06:02 & 12:59:34 \\
2019-03-16 & 11:17:06 & 14:10:40 & & 2019-09-18 & 10:05:07 & 12:58:41 \\
2019-03-19 & 11:26:03 & 14:19:37 & & 2019-09-21 & 10:28:02 & 13:15:18 \\
            &           &        & & 2019-09-24 & 20:51:02 & 23:44:36 \\
\hline
2020-02-19 & 13:10:02 & 16:03:36 & & 2020-08-25 & 11:43:07 & 14:32:46 \\
2020-02-22 & 11:00:09 & 13:53:42 & & 2020-08-28 & 13:59:01 & 16:52:34 \\
2020-02-25 & 10:20:38 & 13:14:12 & & 2020-08-31 & 11:38:05 & 14:31:38 \\
2020-02-28 & 11:06:01 & 13:59:35 & & 2020-09-03 & 11:33:08 & 14:26:42 \\
2020-03-01 & 10:20:37 & 13:14:10 & & 2020-09-06 & 05:49:05 & 08:42:38 \\
2020-03-05 & 10:55:00 & 13:48:34 & & 2020-09-09 & 04:05:59 & 06:59:32 \\
2020-03-08 & 11:08:18 & 13:58:49 & & 2020-09-12 & 02:04:15 & 04:57:49 \\
2020-03-11 & 13:26:02 & 16:14:49 & & 2020-09-15 & 11:27:07 & 14:20:41 \\
2020-03-14 & 10:30:08 & 13:23:42 & & 2020-09-18 & 06:16:36 & 09:10:09 \\
2020-03-17 & 11:25:21 & 14:18:56 & & 2020-09-21 & 10:22:47 & 13:14:48 \\
2020-03-20 & 11:25:03 & 14:14:53 & & 2020-09-24 & 11:17:23 & 14:07:53 \\
\hline
2021-02-23 & 14:40:06 & 17:33:39 & & 2021-08-23 & 10:45:05 & 13:38:39 \\
2021-02-26 & 11:08:08 & 14:01:40 & & 2021-08-26 & 11:22:08 & 14:15:40 \\
2021-03-01 & 11:12:05 & 14:05:39 & & 2021-08-29 & 10:44:38 & 13:38:12 \\
2021-03-04 & 11:19:08 & 14:12:41 & & 2021-09-02 & 11:42:38 & 14:36:13 \\
2021-03-07 & 11:25:07 & 14:18:39 & & 2021-09-04 & 11:16:08 & 14:09:41 \\
2021-03-10 & 11:30:22 & 14:23:55 & & 2021-09-07 & 11:56:22 & 14:49:57 \\
2021-03-13 & 21:21:01 & +1 00:14:34 & & 2021-09-10 & 06:46:31 & 09:39:54 \\
2021-03-16 & 11:38:22 & 14:31:56 & & 2021-09-13 & 11:17:07 & 14:10:41 \\
2021-03-19 & 10:15:22 & 13:08:56 & & 2021-09-16 & 11:25:38 & 14:19:11 \\
2021-03-23 & 11:53:05 & 14:46:38 & & 2021-09-19 & 11:05:07 & 13:58:40 \\
2021-03-25 & 19:36:02 & 22:29:36 & & 2021-09-22 & 12:05:06 & 14:58:40 \\
\hline
\end{longtable}

\begin{figure*}
\centering
\includegraphics[bb=92 273 508 564,clip,angle=0,width=\textwidth]{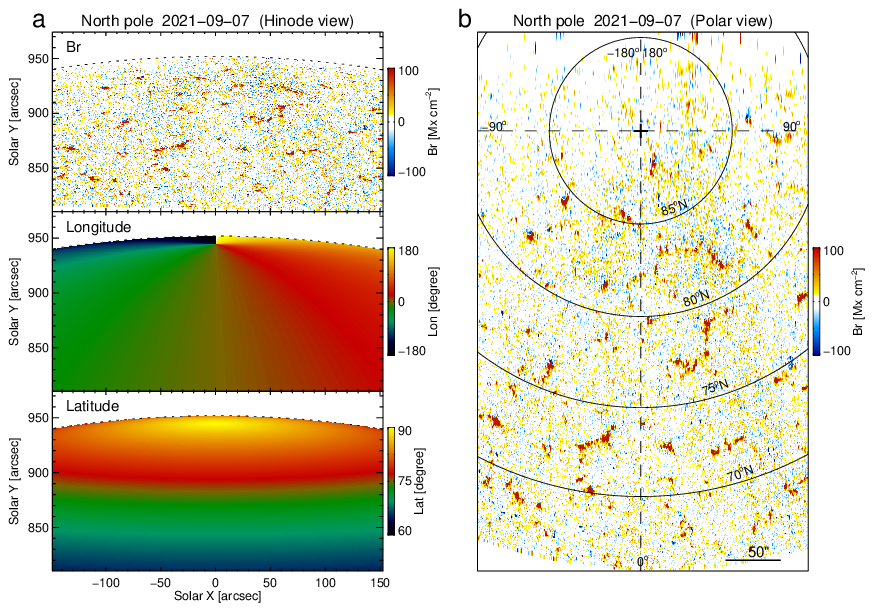}
\caption{Example to illustrate the projection from the Hinode view to the polar view.
(a) Radial magnetic field (top panel) observed by Hinode on 7 September 2021 and corresponding longitude (middle panel) and latitude (bottom panel).
The dotted curves delineate the limb of the solar disk.
(b) Radial magnetic field mapped to be seen from above the north pole, where east is to the left and west is to the right.
The solid curves indicate the latitude separated by 5$^\circ$, the dashed lines indicate the longitude separated by 90$^\circ$, and the plus sign marks the pole.}
\label{fig_LatLon_20210907}
\end{figure*}

\begin{figure*}
\centering
\includegraphics
[bb=92 265 510 588, clip,angle=0,width=\textwidth]{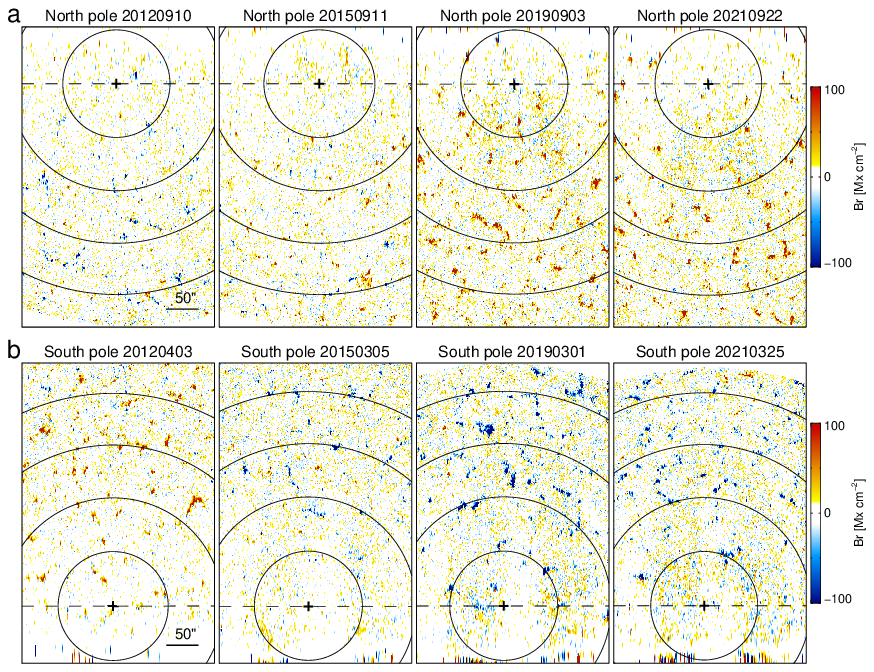}
\caption{Polar view of the radial magnetic flux distribution in the polar caps observed by Hinode.
(a) Radial magnetic field in the north polar cap measured in 2012, 2015, 2019, and 2021.
(b) Similar to (a) but for the south pole. East is to the left and west is to the right.
The plus signs mark the poles, and the solid curves indicate the latitude separated by 5$^\circ$. The dashed lines indicate the boundary, beyond which the region is not considered.}
\label{fig_BrTop_100G}
\end{figure*}

\begin{figure*}
\centering
\includegraphics[bb=160 201 428 637,clip,angle=0,width=0.6\textwidth]{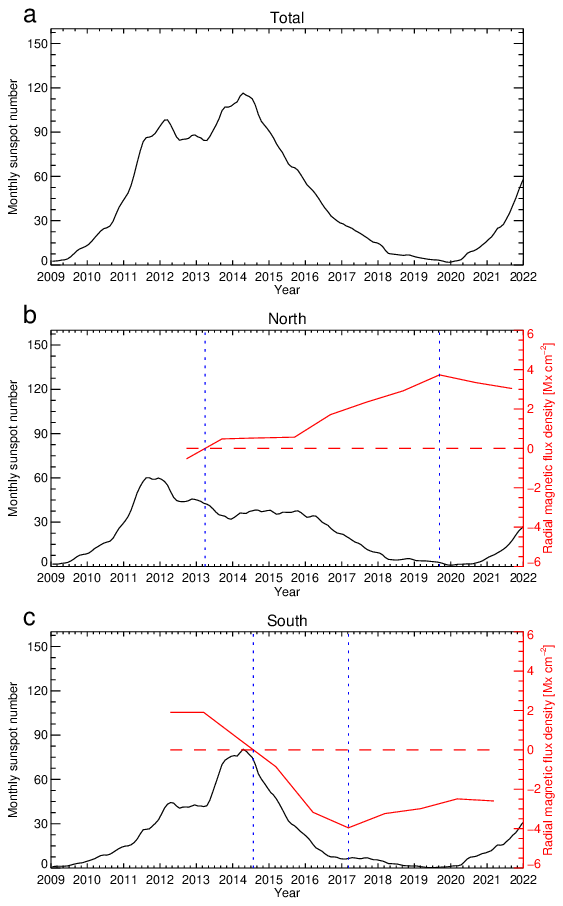}
\caption{The 13-month smoothed monthly sunspot number and the averaged radial magnetic flux density in the latitude range from $\pm$70$^{\circ}$ to the poles. (a) Smoothed monthly sunspot number vs. time from 2009 to 2022. (b) Smoothed monthly sunspot number (black curve) in the northern hemisphere and the averaged radial flux density (red curve) in the north polar cap. (c) Similar to (b) but in the southern hemisphere. Times of the polarity reversal and flux density maximum in each polar cap are marked by the left and right vertical lines, respectively.}
\label{fig_sunspots}
\end{figure*}

\begin{figure*}
\centering
{\subfigure{\includegraphics[bb=186 252 396 586, clip,angle=0,width=0.49\textwidth]{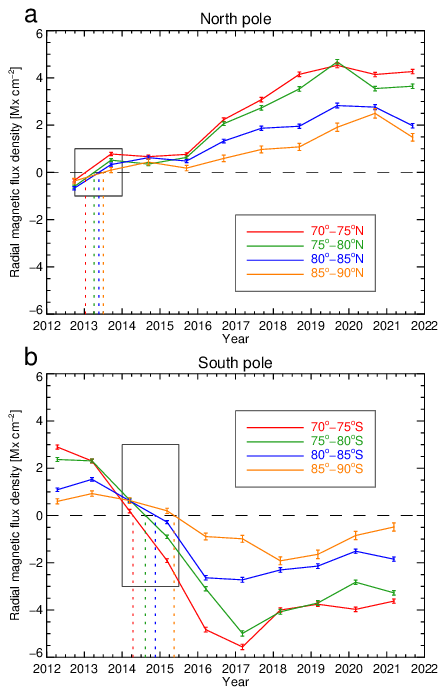}}
\subfigure{\includegraphics[bb=183 252 396 586, clip,angle=0,width=0.49\textwidth]{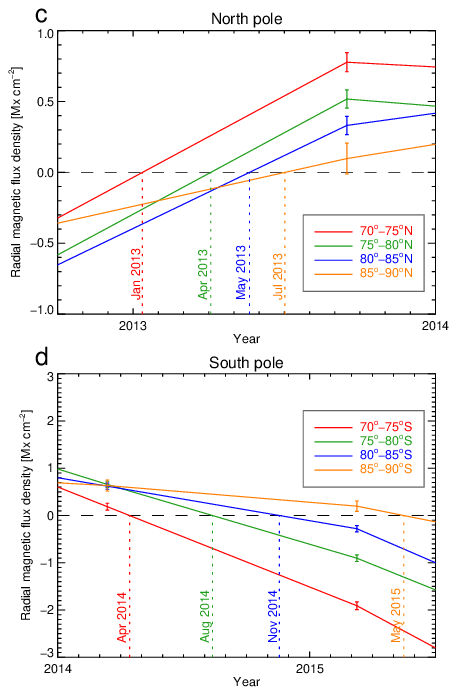}}}
\caption{Long-term variation of the radial magnetic flux density at different latitudes in the north and south polar caps.
(a) Averaged radial magnetic flux density versus time in the north polar cap.
(b) Similar to (a) but in the south polar cap.
The rectangles in (a) and (b) outline the polarity reversal periods as shown in (c) and (d). (c) Variation of the polar magnetic flux density at different latitudes during the polarity reversal process in the north polar cap.
(d) Similar to (c) but in the south polar cap.
The colored curves represent the averaged radial magnetic flux density in different latitude ranges, and the vertical lines mark the times when the polarity reversals were completed.
Each error bar represents 3$\sigma$, where $\sigma$ is the standard error.
\label{fig_BrLat}}
\end{figure*}

Due to the existence of the 7.25$^\circ$ tilt angle of the solar rotation axis with respect to the ecliptic plane, the south and north poles are tilted toward the Earth every March and September, respectively (Petrie 2015).
In order to observe the solar poles at a large viewing angle, Hinode's observation date for the south pole is mainly March, while that for the north pole is September. For each polar cap every year, we use a dataset which consists of $\sim$10 magnetograms with $\sim$3 day interval (see Table \ref{tbl}).
The fields of view of the polar magnetograms cover the latitudes from about $\pm$67$^\circ$ to $\pm$90$^\circ$ and beyond.
For each magnetogram observed from the Hinode view, it can be projected to the top view from above the solar pole, which is needed to correctly show the size distribution and spatial extent of the polar magnetic patches.
Figure \ref{fig_LatLon_20210907} shows an example to illustrate the projection from the Hinode view to the polar view. For a given observed magnetogram, we calculate the longitude and latitude of each pixel. Then the magnetogram is warped to the polar view image via interpolation.

Figure \ref{fig_BrTop_100G} displays the landscape of the radial magnetic fields in the north and south polar caps within ten years.
For the north polar cap (Fig. \ref{fig_BrTop_100G}a), the magnetic field in 2012 was dominated by the negative polarity. Then the magnetic field became weak, the large magnetic concentrations were rare, and both the negative and positive concentrations spread over the polar cap in 2015. While in 2019, the north polar cap was dominated by strong magnetic elements with positive polarity. In 2021, although the dominant field was still the positive polarity, the number of strong positive flux concentrations decreased. On the contrary, the south polar cap (Fig. \ref{fig_BrTop_100G}b) in 2012 was dominated by the positive magnetic elements. Then in 2015, the number and average size of the positive major concentrations decreased and many negative concentrations appeared. As the south polar field had reversed, the strength of the negative polarity field became strong in 2019, and then decreased. In 2021, the dominant positive polarity changed to somewhat weak.

We also use 13-month smoothed monthly hemispheric sunspot number from the Sunspot Index and Long-term Solar Observations (SILSO; online sunspot number catalogue\footnote[2]{https://sidc.be/SILSO/datafiles/}).
The 13-month smoothed monthly sunspot number and the averaged magnetic flux density above 70$^\circ$ latitude are shown in Fig. \ref{fig_sunspots}.
The averaged radial magnetic flux density is calculated as $\sum_{ij}{Br_{ij}S_{ij}}/\sum_{ij}{S_{ij}}$, where $Br_{ij}$ is the radial magnetic flux density and $S_{ij}$ is the corrected real area in each pixel $i$ within the latitude range $\pm$(70--90)$^\circ$ and the longitude range [-90, 90]$^\circ$ in each magnetogram $j$ of the corresponding dataset ($\sim$10 magnetograms) every year for the north and south polar caps. The total sunspot number curve (Fig. \ref{fig_sunspots}a) contains two peaks. The first peak corresponds to the maximum of the sunspot number in the northern hemisphere (Fig. \ref{fig_sunspots}b), and the second one corresponds to that in the southern hemisphere (Fig. \ref{fig_sunspots}c).
In the northern hemisphere, the sunspot number reached the maximum in August 2011. The polarity reversal in the north polar cap took place first, and was completed in March 2013 (Fig. \ref{fig_sunspots}b).
The sunspot number in the southern hemisphere reached the maximum in April 2014, and the polarity reversal in the south polar cap was completed in August 2014, approximately one and a half years after that near the north pole (Fig. \ref{fig_sunspots}c).

After the polarity reversal in the north polar cap, the averaged flux density maintained at a quite weak level of about 0.5 Mx cm$^{-2}$ until the end of 2015.
In the north polar cap, the averaged flux density reached the peak (3.7 Mx cm$^{-2}$) in the end of 2019, i.e., the epoch of the solar minimum (Fig. \ref{fig_sunspots}b). While in the south polar cap, the maximum ($-$4.0 Mx cm$^{-2}$) of the averaged flux density was much earlier, in 2017 (Fig. \ref{fig_sunspots}c).

To investigate the variation of magnetic fields at different latitudes, the radial flux density within every 5$^\circ$ from $\pm$70$^\circ$ latitude to the poles is averaged. Figure \ref{fig_BrLat} shows the long-term variation of the radial magnetic flux density at different latitudes in the north and south polar caps.
The polarity inversion time marked by each vertical line indicates completion of the reversal process in each latitude range. For both the north and south polar caps, the magnetic field in the latitude range $\pm$(70--75)$^\circ$ changed its polarity first, and then the polarity reversal in the higher latitude ranges occurred successively, i.e., the higher latitude corresponds to the later polarity reversal.
In the north polar cap, the reversal times in the latitude ranges 70--75$^\circ$, 75--80$^\circ$, 80--85$^\circ$, and 85--90$^\circ$ are about 2013 January, April, May, and July, respectively (Fig. \ref{fig_BrLat}c). While in the south polar cap, the reversal times in the latitude ranges $-$(70--75)$^\circ$, $-$(75--80)$^\circ$, $-$(80--85)$^\circ$, and $-$(85--90)$^\circ$ are about 2014 April, August, November, and 2015 May, respectively (Fig. \ref{fig_BrLat}d). The polarity reversal in the north and south polar caps lasted for almost half a year and one year, respectively.

Moreover, in the north polar cap, the strongest magnetic flux density in the latitude range 85-90$^\circ$ was observed in 2020, and those in the other lower latitude ranges were observed in 2019 (Fig. \ref{fig_BrLat}a). While in the south polar cap, the strongest magnetic flux density in the latitude range $-$(85-90)$^\circ$ was in 2018, and those in the other lower latitude ranges were in 2017 (Fig. \ref{fig_BrLat}b).

\section{Conclusions and Discussion}

Using the Hinode spectro-polarimetric observations from 2012 to 2021, we investigate the long-term variation of the polar magnetic fields at different latitudes. It is shown that the polarity reversal in the north polar cap took place first, and about one and a half years later, the magnetic polarity in the north polar cap began to reverse.
We find that the magnetic field polarity in each polar cap reversed from the 70$\degr$ latitude to the pole successively at the epoch of solar maximum, i.e., the higher latitude corresponds to the later polarity reversal.

The polar magnetic fields are predominantly unipolar during most of the solar cycle and change the polarity at the epoch of solar maximum (Petrie 2015). The polarity reversal process is usually non-simultaneous in the north and south polar caps (Svalgaard \& Kamide 2013; Pishkalo 2019). For the past solar cycle 24, many studies are dedicated to the polar magnetic field reversals using data from the Wilcox Solar Observatory, the Solar Dynamic Observatory, the National Solar Observatory at Kitt Peak, etc., and as expected, the magnetic field in the north polar region began to reverse first (Svalgaard \& Kamide 2013; Karna et al. 2014; Sun et al. 2015; Janardhan et al. 2018).
Using the Hinode vector magnetograms obtained between 2008 and 2012, the variation of the polar magnetic fields before the polar field reversal during solar cycle 24 was studied by Shiota et al. (2012), and it was found that the net flux in the polar cap decreased in the rising phase, which was more in the north polar cap.
As demonstrated in Fig. \ref{fig_BrTop_100G} and Fig. \ref{fig_sunspots}, the Hinode observations covering the reversal stage show that the polarity reversal in the north polar cap took place first, and then occurred in the south polar cap. This result confirms that the reversal processes in the north and south poles are non-simultaneous, reflecting the solar activity asymmetry between the two hemispheres (Norton \& Gallagher 2010; Upton \& Hathaway 2014).

The reversal in the north polar cap was completed about one year after the maximum of hemispheric sunspot number, and the reversal time of the south pole was at the local cycle maximum in the southern hemisphere. In the north polar cap, the averaged flux density after the polarity reversal maintained at a quite weak level until the end of 2015 (Fig. \ref{fig_sunspots}b and Fig. \ref{fig_BrLat}a). This might be caused by the appearance of non-Joy and anti-Hale active regions (ARs) and the remnant flux surges to the pole (Mordvinov \& Kitchatinov 2019).
The averaged flux density in the north polar cap reached the peak at the epoch of the solar minimum (Fig. \ref{fig_sunspots}b), but the maximum of the averaged flux density in the south polar cap was much earlier, in 2017 (Fig. \ref{fig_sunspots}c). This may be due to the prominent poleward surge during solar cycle 24 in the southern hemisphere caused by AR 12192 (Wang et al. 2020).

The polarity reversal times shown in Fig. \ref{fig_BrLat} revel that the higher latitude corresponds to the later polarity reversal. The time lags among the polarity reversals reflect the magnetic flux migration from the lower latitudes to the poles.
The strongest polar magnetic density appeared earlier than those at the lower latitudes, also implying the existence of the poleward magnetic flux migration in the polar caps.
The remnant magnetic flux from ARs migrates poleward and cancels the polar field of the old solar cycle, thus leading to the polarity reversal in the polar caps (Kaithakkal et al. 2015).

As shown in Figs. \ref{fig_LatLon_20210907} and \ref{fig_BrTop_100G}, the weakest magnetic fields in each north/south polar map are concentrated along the top/bottom edge of the map. It seems that the fall-off of the magnetic field strength near the limb is likely caused by decreasing signal to noise ratio toward the limb. The effects of these signal to noise ratio problems are also visible in the plots of radial magnetic flux density versus time in Fig. \ref{fig_BrLat}, in particular the left panels. Petrie (2022) analyzed equivalent Hinode/SP vector data processed at the High Altitude Observatory, discussed the systematic errors and latitude-dependent changes in some detail (with reference to longitudinal magnetogram data), and demonstrated the reversal of polar field polarity first at lower and then at progressively higher latitudes by constructing polar synoptic maps. The novelty in the present paper (besides the new Nagoya University dataset) is that the magnetic flux densities are computed over separate latitude bands and plotted over time to show the latitude-dependent behavior.

\normalem
\begin{acknowledgements}
We thank Prof. Yukio Katsukawa at NAOJ for the helpful discussion.
This research is supported by the National Key R\&D Programs of China (2019YFA0405000,
2022YFF0503800, 2022YFF0503000), the Strategic Priority Research Programs of the Chinese Academy of Sciences (XDB0560000, XDB41000000), the National Natural Science Foundations of China (12173005, 12273060, 12350004, 12273061, 12222306, and 12073001), the Youth Innovation Promotion Association CAS, and Yunnan Academician Workstation of Wang Jingxiu (No. 202005AF150025).
The data are used courtesy of Hinode and SILSO teams.
Hinode is a Japanese mission developed and launched by ISAS/JAXA, with NAOJ as domestic partner and NASA and STFC (UK) as international partners. It is operated by these agencies in co-operation with ESA and NSC (Norway). ISEE Database for Hinode SOT Polar Magnetic Field (doi: 10.34515/DATA\_HSC-00001) was developed by the Hinode Science Center, Institute for Space-Earth Environmental Research (ISEE), Nagoya University.
The smoothed monthly hemispheric sunspot number data are from the World Data Center SILSO, Royal Observatory of Belgium, Brussels.
\end{acknowledgements}

{}

\clearpage


\begin{thebibliography}{}

\bibitem[Cameron et al.(2016)]{2016ApJ...823L..22C} Cameron, R.~H., Jiang, J., \& Sch{\"u}ssler, M.\ 2016, \apjl, 823, L22
\bibitem[Charbonneau(2020)]{2020LRSP...17....4C} Charbonneau, P.\ 2020, Living Reviews in Solar Physics, 17, 4
\bibitem[Deng et al.(1999)]{1999ScChA..42.1096D} Deng, Y., Wang, J., \& Ai, G.\ 1999, Science in China A: Mathematics, 42, 1096
\bibitem[Homann et al.(1997)]{1997SoPh..175...81H} Homann, T., Kneer, F., \& Makarov, V.~I.\ 1997, \solphys, 175, 81
\bibitem[Ito et al.(2010)]{2010ApJ...719..131I} Ito, H., Tsuneta, S., Shiota, D., et al.\ 2010, \apj, 719, 131
\bibitem[Janardhan et al.(2018)]{2018A&A...618A.148J} Janardhan, P., Fujiki, K., Ingale, M., et al.\ 2018, \aap, 618, A148
\bibitem[Jin \& Wang(2011)]{2011ApJ...732....4J} Jin, C. \& Wang, J.\ 2011, \apj, 732, 4
\bibitem[Kaithakkal et al.(2015)]{2015ApJ...799..139K} Kaithakkal, A.~J., Suematsu, Y., Kubo, M., et al.\ 2015, \apj, 799, 139
\bibitem[Karna et al.(2014)]{2014SoPh..289.3381K} Karna, N., Hess Webber, S.~A., \& Pesnell, W.~D.\ 2014, \solphys, 289, 3381
\bibitem[Kosugi et al.(2007)]{2007SoPh..243....3K} Kosugi, T., Matsuzaki, K., Sakao, T., et al.\ 2007, \solphys, 243, 3
\bibitem[Lin et al.(1994)]{1994SoPh..155..243L} Lin, H., Varsik, J., \& Zirin, H.\ 1994, \solphys, 155, 243
\bibitem[Lites et al.(2013)]{2013SoPh..283..579L} Lites, B.~W., Akin, D.~L., Card, G., et al.\ 2013, \solphys, 283, 579
\bibitem[Mordvinov \& Kitchatinov(2019)]{2019SoPh..294...21M} Mordvinov, A.~V. \& Kitchatinov, L.~L.\ 2019, \solphys, 294, 21
\bibitem[Norton \& Gallagher(2010)]{2010SoPh..261..193N} Norton, A.~A. \& Gallagher, J.~C.\ 2010, \solphys, 261, 193
\bibitem[Orozco Su{\'a}rez \& Del Toro Iniesta(2007)]{2007A&A...462.1137O} Orozco Su{\'a}rez, D. \& Del Toro Iniesta, J.~C.\ 2007, \aap, 462, 1137
\bibitem[Petrie(2015)]{2015LRSP...12....5P} Petrie, G.~J.~D.\ 2015, Living Reviews in Solar Physics, 12, 5
\bibitem[Petrie(2022)]{2022ApJ...941..142P} Petrie, G.~J.~D.\ 2022, \apj, 941, 142
\bibitem[Pishkalo(2019)]{2019SoPh..294..137P} Pishkalo, M.~I.\ 2019, \solphys, 294, 137
\bibitem[Shiota et al.(2012)]{2012ApJ...753..157S} Shiota, D., Tsuneta, S., Shimojo, M., et al.\ 2012, \apj, 753, 157
\bibitem[Sun et al.(2015)]{2015ApJ...798..114S} Sun, X., Hoeksema, J.~T., Liu, Y., et al.\ 2015, \apj, 798, 114
\bibitem[Svalgaard \& Kamide(2013)]{2013ApJ...763...23S} Svalgaard, L. \& Kamide, Y.\ 2013, \apj, 763, 23
\bibitem[Tang \& Wang(1991)]{1991SoPh..132..247T} Tang, F. \& Wang, H.\ 1991, \solphys, 132, 247
\bibitem[Tsuneta et al.(2008)]{2008ApJ...688.1374T} Tsuneta, S., Ichimoto, K., Katsukawa, Y., et al.\ 2008a, \apj, 688, 1374
\bibitem[Tsuneta et al.(2008)]{2008SoPh..249..167T} Tsuneta, S., Ichimoto, K., Katsukawa, Y., et al.\ 2008b, \solphys, 249, 167
\bibitem[Upton \& Hathaway(2014)]{2014ApJ...780....5U} Upton, L. \& Hathaway, D.~H.\ 2014, \apj, 780, 5
\bibitem[Wang et al.(2020)]{2020ApJ...904...62W} Wang, Z.-F., Jiang, J., Zhang, J., et al.\ 2020, \apj, 904, 62

\end{thebibliography}
\end{document}